# Measuring research data reuse in scholarly publications using generative artificial intelligence: Open Science Indicator development and preliminary results

Lauren Cadwallader*, Iain Hrynaszkiewicz*, parth sarin** and Tim Vines**

* *lcadwallader@plos.org*; *ihrynaszkiewicz@plos.org*

*0000-0002-7571-3502; 0000-0002-9673-5559*

*PLOS, Cambridge, UK*

***parth@dataseer.ai; tim@dataseer.ai*

*0009-0004-9317-1446;0000-0002-5418-9967*

*DataSeer, Vancouver, Canada*

**Abstract**

Numerous metascience studies and other initiatives have begun to monitor the prevalence of open science practices when it is more important to understand the "downstream" effects or impacts of open science. PLOS and DataSeer have developed a new LLM-based indicator to measure an important effect of open science: the reuse of research data. Our results show a data reuse rate of 43%, which is higher than established bibliometric techniques. We show that data reuse can be measured at scale using LLMs and generative artificial intelligence. The positive effects of research data sharing and reuse may currently be underestimated.

## 1. Introduction

Open and responsible research practices are assumed to provide benefits to the research enterprise, and to society. These benefits, which have been evidenced to varying extents, include increased trust in research, increased collaboration, increased efficiency, greater potential for verification and replication of research, as well as societal impacts (Cole et al. 2024; Klebel et al. 2025; Tsipouri et al. 2025) – enabled by greater transparency in research outputs and processes, and greater use and reuse of research outputs, such as articles, data, code, materials, and methods. However, open science policies and practices require financial and other kinds of support to implement effectively, and can be viewed as an investment in better research outcomes more than a cost. It is nevertheless important for policy makers, funders and others implementing policies or seeking to understand researcher practices to understand if open science policies lead to positive, or unintended, impacts on research. Greater attention on open science policies and advances in research methodologies, supported by automation and artificial intelligence have enabled the prevalence of open science practices to be measured and monitored. Open science policies, such as the French National Open Science Plan and PLOS' data availability policy, through open science monitoring, are demonstrably associated with increasing prevalence of open science practices – such as open access publishing, data sharing, and code and software sharing (Colavizza et al. 2020; Colavizza et al. 2025). In 2024, the Open Science Monitoring Initiative (OSMI) launched, in recognition of increasing global attention on monitoring and open science, linked to worldwide adoption of the UNESCO recommendation open science (UNESCO 2021). While increasing prevalence of open practices is valuable – and is a prerequisite to them having any effects – understanding the impact of open science and associated policies is more important as impact measures are closer to demonstrating or understanding *why* particular practices should be promoted.

*1.1 Previous methods to measure reuse of data*

There is a growing attention on understanding the impacts of research datasets and using measurable indicators to do so. To date, research data reuse has generally been investigated via surveys and interviews with researchers. One survey, from members of our group, of 617 researchers across research disciplines, career stages and geographies found that 52% of researchers reported reusing research data (Hrynaszkiewicz, Harney & Cadwallader 2021). A much larger survey of authors in 20 disciplines (n=1,769) reported data reuse at 54.3% (Khan, Telwall & Kousha 2023). Reuse of research results and ideas are routinely measured through citations and bibliometrics. In principle, bibliometric approaches can be used to measure data reuse, since the citation of datasets deposited in repositories that assign persistent identifiers, such as DOIs, enables data to be cited in reference lists (Cousijn et al. 2018). However, despite widespread support of data citation by publishers, infrastructure providers and others (Stall et al. 2023), it remains at best unevenly implemented by publishers and infrequently adopted by researchers. Moreover, some of the most commonly shared and reused datasets, such as genetic sequences, are not assigned persistent identifiers (Datacite & Make Data Count 2025), making standard bibliometric citation infeasible. Indeed, bibliometric approaches to measuring data citations have implied a reuse rate of 9-16% (while not accounting for data citations for a purpose other than reuse). In another study describing a bespoke workflow for detecting data sharing reuse reported a reuse rate of 1.8% (Iarkaeva, Nachev & Bobrov 2024). Further, of nearly 2 billion citation links in Crossref, only a few hundred thousand are tagged as dataset citations. All of these observations point to a significant gap between data reuse as detected by current methods and the rates implied by statements from researchers themselves.

The UK Reproducibility Network (UKRN) is a peer-led consortium that promotes rigorous research practices, including open research. In 2023-2025, UKRN ran several pilots to develop and test open research indicators that were identified as priorities by UK research institutions, including a pilot to measure the effects (impacts) of an important open research practice, data sharing – or determine if they are measurable (Jacobs et al. 2026). Since 2022, the open science publisher, policy and practice organisation PLOS has produced its own Open Science Indicators (OSI) to measure the prevalence of open science practices, in collaboration with DataSeer, and intended to measure reuse of research data as part of its plans (Hrynaszkiewicz & Kiermer 2022). PLOS and DataSeer together collaborated with UKRN and 15 UK-based research institutions, plus other indicator providers, to develop new concepts and potential solutions to measure the effects of data sharing. The UKRN indicator pilots served as a mechanism for PLOS and DataSeer to develop their data reuse indicator, providing mutual benefits to both initiatives. Measuring data reuse and reusability is also of general interest because data generation and sharing are the most prevalent practices measured by PLOS OSI, and data reuse (including FAIRness) is of interest to funders, policymakers, publishers, and institutions that in the context of growing numbers research data management and sharing policies from these stakeholders (Sellanga et al. 2024). Reuse of research data and databases has been associated with economic and other benefits in numerous case studies (Rosemberg et al. 2026), but few studies have attempted to quantify the prevalence and impact of data reuse through content analysis, across a broad segment of researchers. Open science practices, such as data sharing, have been linked to greater academic impact through increased citations, and data reuse has been linked to enhanced citations (Yang et al. 2025).

More recently, Yang and Colavizza (2025) utilised a novel automated using manual annotation and a fine-tuned SciBERT model to detect mentions of shared data in publications and then classified those mentions into three categories: release, reuse, reference. Approximately 30% of datasets were classified as an intention to reuse data in this study, much higher than bibliometric approaches. Some differences in results may be accounted for by differences in definitions and measures of data reuse between studies, but given the differences between what researchers report in qualitative studies and what is detected via automation, we assumed that the full extent of data reuse in published research is not well known at scale. Moreover, we began our investigation by first, in a technology agnostic fashion, defining data reuse based on a literature review and community consultation. Importantly, this definition includes measuring not just the prevalence of data reuse but additional context as to how or why data have been reused.

## 2. Methods

### *2.1 Defining a reuse indicator*

We first conducted a literature review (Borgman & Groth 2024; Gregory et al. 2023; Lafia et al. 2023; van de Sandt et al. 2019; Wallis, Rolando & Borgman 2013; Zuiderwijk, Shinde & Jeng 2020) and then a consultation of institutions, funders, meta-researchers to establish the scope and requirements for an indicator of data reuse. The collaboratively developed requirements for a data reuse indicator were shared openly for public comment and consultation including with UKRN pilot institutions. The top-level requirements are:

1. Identify if a research article reuses research data
2. Identify which dataset(s) have been reused in the article
3. Identify how or why each dataset was reused by the article's authors

Following our public consultation, data reuse was initially defined as the analysis of existing data collected by other individuals, institutions or by the same researcher for a new research purpose, which has been published in a research article. This definition was intentionally broad and included secondary analysis, reanalysis, replication, and reproducing previous results as types of "new research" that could be derived from reusing data. The prototype dataset was designed to include information about whether the new research was a meta-analysis or systematic review or neither. Data generation was defined as "any research data (information collected, observed, or created for purposes of producing original research results) supporting the results of published article" following the data generation definition in the original development of the Open Science Indicators (Hrynaszkiewicz & Kiermer 2022).

*2.2 Developing a large language model (LLM)*

To meet these indicator requirements, the project team initially developed a language model by fine-tuning Llama 3.1 8B on the task of identifying data reuse and sharing and analysing 4328 PLOS articles. This produced a prototype data output that was tested with researchers, librarians (including the UKRN pilot institutions), policy makers, and journal editors. Through this user testing it was validated that a modified version of the LLM's reasoning trace can meet the third, most challenging, requirement. Following this user testing the definition of data reuse was also modified to exclude systematic reviews from the definition, based on feedback from users. More detailed methods for the development of this earlier prototype are available in Appendix 4 of Jacobs et al. (2026). The prototype dataset produced using this method is available in PLOS & DataSeer (2026).

For this project, we fine-tuned a large language model to identify whether an article reports new data generation and/or data reuse and to extract evidence supporting those judgments. We prompted the model with the markdown version of an article text and designed it to output a structured JSON object containing boolean indicators for generation/reuse, lists of citations, accession numbers, URLs, and DOIs associated with each category, and free-text descriptions summarizing the data generation and reuse practices in the article.

Our training and test dataset consisted of 1,732 PLOS articles, curated by the DataSeer annotation team. For each article, annotators labeled whether data were generated and/or reused, and identified relevant citations, accession numbers, URLs, and DOIs. To improve label quality, we performed a model-assisted verification pass using Qwen3-32B. We flagged 296 articles for manual reexamination; curators corrected 132, and 164 were left unchanged. To improve robustness, each article was converted to markdown using two pipelines: markitdown (Microsoft 2026) and Docling (Auer et al. 2024). We also truncated articles which resulted in prompts longer than 25,000 tokens by adding a [TRUNCATED] marker in the middle, preserving the beginning and end of the article (generally leaving the abstract, data availability statement, and references intact).

We used Qwen3-32B as the base model and fine-tuned it with LoRA adapters (Hu et al. 2021) applied to all of the linear layers, except the final language modeling head. Training was conducted on 8 NVIDIA H200 GPUs with DeepSpeed in fp16. Training consisted of two stages: self-distillation and reinforcement learning with Group Relative Policy Optimization (GRPO). For the supervised stage, we scored the rollouts from the verification run and fine-tuned on high reward ($\geq 0.9$) completions, training on a total of 27.6M tokens. We then refined the model using GRPO initialized from the supervised model for 1 epoch.

To guide reinforcement learning, we designed a reward function to enforce three properties: outputs had to be validly structured, grounded in the source article, and consistent with the target annotation. The overall reward function was defined as

$$r(p,\ c,\ y)\ =f(c)\times e(p,\ c)\times v(c,\ y),$$

where $p$ is the article text, $c$ is the model completion, and $y$ is the target annotation. Each component is in $[0,\ 1]$ so the multiplicative formulation assigns a low reward to completions that failed on any one criteria. The first component, format reward $f(c)$, was binary and rewarded completions that could be parsed correctly. The second component, "embellishment" reward $e(p,\ c)$, penalized predictions that introduced unsupported or hallucinated string values. And the third component, accuracy reward $v(c,\ y)$, measured agreement with the target annotation, allowing for partial credit.

To avoid contradictory optimization signals, we also filtered the ground truth using these same fuzzy-matching thresholds, removing examples where markdown conversion noise might prevent the model from achieving high reward. This removed approximately 68 articles whose annotated evidence did not appear in the article text. We evaluated the model on a held-out test set of 296 articles, taking 3 samples per article, and the accuracy on the boolean fields achieved by the model compared to the DataSeer annotators is given in Table 1 and the F1 scores achieved by the model given in Table 2.

Table 1: Accuracy of the LLM compared to DataSeer annotators for the boolean fields.

| **Field** | **pass@1** | **pass@3** |
|---|---|---|
| new_data_generated | 89.8% | 93.2% |
| reuse_data | 79.4% | 85.5% |

Table 2: The F1 scores achieved by the model on each of the fields.

| **Field** | **pass@1** | **pass@3** |
|---|---|---|
| new_data_citations | 0.997 | 1.000 |
| new_data_accessions | 0.953 | 0.964 |
| new_data_dois | 0.957 | 0.979 |

| new_data_urls | 0.859 | 0.889 |
|---|---|---|
| reuse_data_dois | 0.929 | 0.956 |
| reuse_data_accessions | 0.865 | 0.876 |
| reuse_data_urls | 0.764 | 0.793 |
| reuse_data_citations | 0.753 | 0.805 |

Although the accuracy for reuse_data is lower than that for new_data_generated (Table 1), the indicator still has high precision (0.866) but tends to make false-negative errors. The relatively high rate of false negatives indicates that the true rate of reuse is almost certainly higher than we report here. The results presented here should be understood as a meaningful step towards better detecting reuse, but there are certainly indications of reuse that remain outside the scope of what we can currently detect. 90% of false negatives are articles where the model correctly flags new data generation but misses concurrent reuse, for example, when a public reference genome or sequence database is incorporated into the authors' own sequencing pipeline but is not named as reuse or benchmarking new data against a previously released dataset is filed as a mere citation. The moderate F1 scores for reuse-related text fields stem from high precision, low recall, i.e., under-extraction not spurious extraction.

Given the ambiguity in evaluating F1 scores on text fields and that our metrics are sensitive to noise in the markdown extraction, we felt this performance was good enough to proceed with a pass@3 verifier approach. This model also outperforms every large open model that we tested, including GLM-5, gpt-oss-120b, Qwen3-235B-A22B, and Deepseek-R1. At inference time, we sampled three completions per article in Q1 2024 (a total of 4,475 articles) and then passed the model responses and article text to a verifier model (the base Qwen3-32B model) which ranked the completions. Our results are based on the highest-ranked completion for each article published by PLOS in Q1 2024.

**3. Results**

The dataset includes the analysis of 4475 research articles published by PLOS between 1st January 2024 and 31st March 2024. Table 3 details the number and percentage of articles generating new data and/or reusing existing data. 59% of articles generated new data and 43% reused data (articles may generate new data, reuse data, or both) and 10% did not generate or reuse data. Of the articles using data, 11% both generated new data as well as reusing data, meaning that the majority of articles either just generate new data or reuse existing data.

Table 3: Data generation and reuse rates according to the LLM output for PLOS research articles published between 1st January 2024 and 31st March 2024. Note, articles may both generate new data and reuse existing data.

| | count | % |
|---|---|---|
| publications | 4475 | |
| new data generated | 2613 | 59% |
| data reused | 1899 | 43% |
| no new data or reuse | 435 | 10% |

Breaking down the reuse data by region (Table 4), we see that the articles only reusing data in Africa are lower than other regions (31%), whereas they are much higher in Asia-Pacific 46%). The rates of data reuse and generation in North America, Latin America and Europe are very similar to each other and fall in between the rates seen in Africa and Asia-Pacific.

Table 4: Data reuse and generation rates by region. Note, region was not assigned to 6 articles. Articles may both generate new data and reuse existing data.

| | North America | | Latin America | | Europe | | Asia-Pacific | | Africa | |
|---|---|---|---|---|---|---|---|---|---|---|
| | Count | % | Count | % | Count | % | Count | % | Count | % |
| publications | 1007 | | 212 | | 1074 | | 1841 | | 335 | |
| new data generated | 639 | 63% | 138 | 65% | 682 | 64% | 926 | 50% | 223 | 66% |
| data reused | 429 | 42% | 88 | 41% | 416 | 39% | 859 | 46% | 104 | 31% |
| no new data or reuse | 74 | 7% | 18 | 8% | 102 | 9% | 209 | 11% | 32 | 10% |

New data generation and data reuse vary considerably by discipline. Table 5 shows the rates for the four broad topic fields (Health Sciences, Life Sciences, Physical Sciences and Social Sciences). Physical Sciences and Social Sciences are notable as they have equal (or near equal) levels of data generation and data reuse. Physical Sciences also has a much higher rate of articles that do not generate new data nor reuse existing data (18%). Physical Sciences and Social Sciences also see the highest amounts of data reuse (47% and 50% respectively). Life Sciences see the highest amount of new data generation (76%) and lowest amount of data reuse (37%).

Table 5: Rates of data reuse and generation by discipline. Notes, articles may both generate new data and reuse existing data.

| | publications | new data generated | | data reused | | no new data or reuse | |
|---|---|---|---|---|---|---|---|
| | | Count | % | Count | % | Count | % |
| Health Sciences | 1842 | 1070 | 58% | 739 | 40% | 185 | 10% |
| Life Sciences | 1029 | 780 | 76% | 382 | 37% | 48 | 5% |
| Physical Sciences | 783 | 368 | 47% | 365 | 47% | 143 | 18% |
| Social Sciences | 817 | 395 | 48% | 409 | 50% | 59 | 7% |

At least one reasoning trace was provided for 98% of the articles (n=4370). 89% of articles (n=4001) had a reasoning trace for whether or not new data generation was detected. 79% of articles (n=3531) had a reasoning trace for data reuse. The LLM output also provides details on DOIs, accessions, URLs and citations associated with generated or reused data. As an example, the LLM output identifies 487 articles (10.9%) as either generating or reusing data with an accession number. A wide range of accession numbers has also been identified using the LLM technique. 1396 unique "accession numbers" are listed in the reused data categories.

## 4. Discussion

As a paper-in-progress reporting an LLM-driven approach still in development, we provide only minimal interpretation of our results to date.

Overall we observe levels of data reuse, as determined by this LLM approach in our sample of 4475 PLOS publications, of 43%. This result is closer to the estimates of data reuse provided by researchers in qualitative research, which typically find more than 50% of researchers self-reporting data reuse – while acknowledging that our unit of analysis is data reuse in publications, rather than data reuse in general. The results from this sample do, however, suggest that bibliometric and other less advanced techniques for measuring data reuse in scholarly publications underestimate data reuse. If validated in a larger sample of publications – which our group is currently working on – these preliminary results suggest that data reuse, and therefore positive effects of research data sharing and open science practices, may currently be underestimated. The results in our sample also point to regional and disciplinary differences that may warrant further exploration. For example, the lower prevalence of data reuse and higher rates of data generation in Africa relative to other regions may be noteworthy in the context of data sovereignty and concerns about "helicopter science" (Haelewaters, Hofmann & Romero-Olivares 2021) .

Compared to the PLOS Open Science Indicators (OSIs) (PLOS 2022), which uses a natural language processing approach to detect data generation and sharing (see DataSeer n.d. for methods), the LLM-based approach reported here gives a more nuanced view of data as a scholarly output. Firstly, the PLOS OSIs used a much simpler definition of data generation - essentially if an article is based on data then data generation was marked as true - which resulted in a very high data generation rate of 98% for the same set of articles published by PLOS in the first quarter of 2024. The LLM-based approach is intended to differentiate between the ways that data is used/created resulting in a lower rate of new data generation (59%). In addition, the LLM-based approach identifies a higher number of articles that do not use data (whether new or existing) in their research (10% compared to 2% in the PLOS OSIs) due to the definition of data generation used, which has been further refined since the UKRN pilots. For the approach reported here, data generation involves taking information from the physical world and moving it onto a computer. Studies involving only simulations or assessments of workflows would not count as generating data nor would systematic reviews or similar kinds of studies that assemble a dataset of values extracted from published studies count as data reuse. This illustrates that the definitions used are crucial for explaining the variance in results and why community alignment on monitoring solutions is vital.

The number of individual data outputs identified by the LLM-based approach is greater than seen in the PLOS OSIs and the structure of the text strings identified by the LLM suggests this is because researchers are describing the data they used (new or existing) in ways that do not conform to widely accepted citation practices (Data Citation Synthesis Group 2014). The problem of poor data citation is an issue that has been discussed (Peters et al. 2016) and the results of our study add to existing knowledge of how data may be mentioned in an article but not as a formal citation (Gregory et al. 2023).

This approach also provides potential insights on the impact and effects data reuse that measuring the prevalence of open science practices alone does not demonstrate. Greater evidence on the impact of research sharing and reuse practices, particularly in the context of organisations with open science and research sharing policies, will be of interest to research funders, policy makers, and other stakeholders. Some funding agencies, for example, such as the Michael J Fox Foundation, and Aligning Science Across Parkinson's are already exploring the potential of this technique (DataSeer 2026).

### *4.1 Limitations*

The limitations of this work include the difficulty in creating a definition of data use versus data reuse as well as the ability to differentiate between reuse of one’s own data vs. someone else’s data. As with any monitoring solution design choices have to be made in how definitions are constructed and applied to the corpus of articles under analysis. Whilst we have endeavoured to consult with the community to create definitions that are supported, there will always be debate as to where to draw particular distinctions between different types of data (re)use. Furthermore, the development of this approach has focused on a limited sample of PLOS articles and we may find different levels of performance when run with a different cohort of articles.

**Open science practices**

The LLM output data analysing the PLOS articles is openly available at https://doi.org/10.6084/m9.figshare.31872850, along with the preliminary analysis of articles conducted as part of the UKRN pilot initiative mentioned in the text. The LLM LoRAs and accompanying code are proprietary to DataSeer (and hence cannot be made public) but enquiries should be directed to dataseer@dataseer.ai.

**Acknowledgements**

We would like to thank Marcel LaFlamme for contributing to the development of the reuse definitions, Souad McIntosh and Alyssa Yong for generating the training data, and Scott Kerr for supporting data processing

**Author contributions**

LC - data curation, formal analysis, writing - original draft, writing - review & editing
IH - conceptualisation, project administration, supervision, writing - original draft
ps - data curation, formal analysis, investigation, methodology, software, validation, writing - original draft
TV - conceptualisation, formal analysis, supervision, writing - original draft

**Competing interests**

IH and LC are employees of PLOS. IH is an initiator and member of the Open Science Monitoring Initiative (OSMI) and a co-author of the Principles of Open Science Monitoring. ps is an Affiliate Researcher at the Public Knowledge Project and the Scholarly Communication Lab, and an organizer of the Collaborative Metadata (COMET) project. No external funding was received for this work. TV is the majority shareholder in DataSeer.